\pdfoutput=1

\documentclass[11pt]{article}
\usepackage{jheppub}

\usepackage{amsmath}
\usepackage{amssymb}
\usepackage{graphicx,color,slashed,braket}
\usepackage{bbm}
\usepackage{ifpdf}
\usepackage{comment}
\usepackage{soul}

\usepackage{booktabs}
\usepackage{multirow}
\usepackage{makecell}
\usepackage{float}
\usepackage{verbatim}
\usepackage{caption}
\usepackage{cancel}
\usepackage{enumerate}
\usepackage{tcolorbox}
\usepackage[font=small,labelfont=bf]{caption}

\usepackage{color}
\definecolor{dark-gray}{gray}{0.20}
\definecolor{gray}{gray}{0.30}
\definecolor{light-gray}{gray}{0.80}
\definecolor{dark-red}{rgb}{0.7,0,0}
\definecolor{dark-green}{rgb}{0.1,0.4,0}
\definecolor{dark-blue}{rgb}{0.3,0.3,0.7}
\definecolor{light-blue}{rgb}{0.8,0.8,1}

\usepackage{tikz}
\usetikzlibrary{calc}

\usepackage{hyperref}
\usepackage{setspace}
\hypersetup{
	colorlinks=true,
	linkcolor=dark-blue,
	citecolor=dark-red,
	urlcolor=dark-green,
	linktoc=page
}
\newcommand{\be}{\begin{equation}}
\newcommand{\ee}{\end{equation}}

\def\be{\begin{equation}}
\def\ee{\end{equation}}
\def\bea{\begin{eqnarray}}
\def\eea{\end{eqnarray}}

\newcommand{\e}{\mathrm{e}}

\newcommand{\dd}{\mathrm{d}}

\newcommand{\vol}{\text{vol}}

\renewcommand{\Im}{\text{Im}\,}
\renewcommand{\Re}{\text{Re}\,}
\renewcommand{\d}{\textrm{d}}

\allowdisplaybreaks

\title{Weak $G_2$-manifolds and scale separation in M-theory from type IIA backgrounds}

\preprint{UUITP-25/24}

\author{Vincent Van Hemelryck}

\affiliation{Department of Physics and Astronomy, Uppsala University, Box 516, SE-75120, Uppsala, Sweden}

\emailAdd{vincent.vanhemelryck@physics.uu.se}

\notoc

\abstract{
This work provides evidence for the existence of supersymmetric and scale-separated AdS$_4$ vacua in M-theory of the Freund-Rubin type. The internal space has weak $G_2$-holonomy, which is obtained from the lift of AdS vacua in massless type IIA on a specific SU(3)-structure with O6-planes.
Such lifts require a local treatment of the O6-planes, therefore going beyond the usual smeared approximation. The setup is analysed by solving the pure spinor equations and the Bianchi identities perturbatively in a small backreaction parameter, preserving supersymmetry manifestly and therefore extending on previous work. This approach is applicable to lifts of other type IIA vacua on half-flat SU(3)-structures, including those with D6-brane sources.
The resulting 7d manifold presented here exhibits singularities originating from the O6-planes loci in type IIA theory. Additionally, scale separation in M-theory arises from a decoupling between the Ricci curvature and the first eigenvalue of the Laplacian of the proposed 7d manifold, thereby challenging certain conjectures in the swampland program.

}
\begin{document}

\maketitle

\newpage
\tableofcontents
\newpage
\section{Introduction}
\label{sec:Introduction}

Given the inherently ten-dimensional nature of critical string theory, the most studied approach in obtaining lower-dimensional theories from string theory is through compactifications. However, to achieve realistic phenomenological models, it is essential to establish a hierarchy between the cosmological scale $L_H$ and the Kaluza-Klein (KK) scale $L_\text{KK}$ of the extra dimensions. Only with this hierarchy, $L_\text{KK} \ll L_H$, the effective field theory can be considered genuinely lower-dimensional.
Yet, the fate of scale-separated vacua is still under a lot of debate in the swampland program with arguments against them, formalised in the strong version of the Anti-de Sitter conjecture
\cite{Lust:2019zwm}. Additionally, it has been shown that for 4d theories with extended supersymmetry, scale separation is in tension with the magnetic weak gravity conjecture, which has been argued to persist in higher-dimensional theories with more than four supercharges \cite{Cribiori:2022trc}. Constraints on scale separation for such theories have been found in a holographic context as well, see refs.
\cite{Bobev:2023dwx,Perlmutter:2024noo} for recent progress. 

Nevertheless, there are limited examples where this feature seems realised, see ref.~\cite{Coudarchet:2023mfs} where the state-of-the-art is reviewed. All those examples enjoy $\mathcal{N}\leq 1$ supersymmetry.  However, it is fair to say that today, there is no consensus on the complete validity of these constructions, see \cite{Coudarchet:2023mfs} and references therein.

In type IIA string theory, which is the focus of this work, such solutions have been studied with Romans mass, the so-called DGKT construction \cite{DeWolfe:2005uu,Camara:2005dc}. To stabilise the moduli classically, one needs NSNS- and  RR fluxes and (intersecting) O6-planes. These models enjoy an unbounded $F_4$-flux, and in the large flux regime, the solutions are scale-separated. Additionally, the situation for AdS$_3$ has been studied in refs.~\cite{Farakos:2020phe,VanHemelryck:2022ynr,Emelin:2021gzx,Emelin:2022cac,Farakos:2023nms,Farakos:2023wps,Arboleya:2024vnp}, where scale-separated solutions appear similarly in type IIA compactifications on manifolds with $G_2$-holonomy \cite{Farakos:2020phe,VanHemelryck:2022ynr}.
However, these works have been scrutinised on several fronts. First, there are intersecting O6-planes that are treated as smeared instead of local, codimension-three objects.
In refs.~\cite{Junghans:2020acz,Marchesano:2020qvg} it was shown that one can formulate a local solution where the local O6-backreaction is encoded in perturbative corrections. The equations were only solved at first order, where the full solution is a linear superposition of single-source solutions, which is insensitive to the intersection of the O-planes. Nevertheless, in ref.~\cite{Emelin:2024vug}, some investigations were performed on potentially dangerous singularities at second order in the perturbation theory but they were found to be absent.\footnote{See also ref.~\cite{Andriot:2023fss} for a study of  corrections to extensions of the DGKT scenario.}
Moreover, the fact that the O6-planes are intersecting in the first place, has also been scrutinised. Although the O-planes are indeed intersecting in manifolds with orbifold singularities such as $\mathbb{T}^6$ orbifolds, the same does not need to be true for smooth Calabi-Yau manifolds. This was shown in ref.~\cite{Junghans:2023yue} where blow-ups of those singularities were considered on $\mathbb{T}^6/\mathbb{Z}_3$ and $\mathbb{T}^6/(\mathbb{Z}_3\times \mathbb{Z}_3)$.

Additionally, it must be emphasised that finding fully local solutions for intersecting D-brane or O-plane sources, i.e. without partially smearing some of the sources, is generally challenging. For example, it is currently not known how such a solution should work for two intersecting D-branes in 10d space with flat asymptotics \cite{Bardzell:2024anh}. However, in setups with sources that are completely intersecting, i.e. with no common transverse directions among the sources, such issues do not seem to arise, see ref.~\cite{Gauntlett:1996pb} for an example.

Even though a local treatment of the O-planes in the DGKT setting has been realised to a certain extent, they still render singularities at their O-plane loci. It is not well-understood whether these singularities are of the allowed type in string theory \cite{DeLuca:2021mcj}.

To avoid some of these issues, ref.~\cite{Cribiori:2021djm} considered a scale-separated setup in massless type IIA string theory, with the opportunity to lift solutions to M-theory. The benefit of such M-theory lifts is that O6-planes lift into the geometry such that there are no local sources.\footnote{See also ref.~\cite{DeLuca:2022wfq} where a non-supersymmetric, scale-separated AdS$_4 \times \mathbb{T}^7$ solution of M-theory was constructed with Casimir energy and fluxes to stabilise the moduli, without the need of local M-brane sources.}
In ref.~\cite{Cribiori:2021djm}, the type IIA theory was compactified on a nilmanifold with negative curvature, the so-called Iwasawa manifold, which is a twisted $\mathbb{T}^6/(\mathbb{Z}_2\times \mathbb{Z}_2)$ orbifold. The setup can be regarded as some double T-dual version of a DGKT setup on a $\mathbb{T}^6/(\mathbb{Z}_2\times \mathbb{Z}_2)$ Calabi-Yau manifold. There are two unbounded $F_2$-fluxes and unbounded $F_6$-flux, and in the large flux regime, the model features scale separation. The manifold needs to be anisotropic, which is accommodated by fluxes that are not scaling similarly. This allows both weakly and strongly coupled solutions, the latter being suitable for M-theory lifts.\footnote{In ref.~\cite{Carrasco:2023hta}, the type IIA construction was generalised to T-duals from massive IIA setups compactified on elliptically fibered Calabi-Yau three-folds.}\footnote{See ref.~\cite{Tringas:2023vzn} for an anisotropic treatment of the DGKT scenario.}
In ref.~\cite{Cribiori:2021djm}, a local perturbative solution was constructed with the techniques of ref.~\cite{Junghans:2020acz}, solving the 10d equations of motion order by order. This local solution was then lifted to M-theory. 
The proposed M-theory solution was also found to be scale-separated, by estimating the KK scale as the largest radius of the 7d manifold. The solution also respects the theorem in ref.~\cite{Gautason:2015tig}, saying that the 7d and the 4d curvature scale (the AdS) scale must be of the same order, not allowing for scale separation if the 7d KK scale is of the same order as the curvature scale. It was found in ref.~\cite{Cribiori:2021djm} that the two latter scales indeed decouple, such that scale separation occurs. Note that this goes against the conjecture formulated in ref.~\cite{Collins:2022nux}. However, a more detailed study of this proposed manifold was not performed.

Nevertheless, it has to be emphasised that the M-theory solution that was constructed with this local type IIA solution, is of the Freund-Rubin type, at least to first order in the perturbative scheme. This means that the M-theory setup comprises only electric $G_4$-flux (spacetime-filling) or magnetic $G_7$-flux (the Hodge dual statement) and a 7d manifold with a positive Einstein metric. 

Now, it is well known that manifolds with a $G_2$-structure and weak $G_2$-holonomy provide such Freund-Rubin solutions \cite{Behrndt:2003uq,Behrndt:2003zg,Lukas:2004ip,Acharya:2003ii,Behrndt:2005im}, partially reviewed in ref.~\cite{Koerber:2010bx,Tomasiello:2022dwe}. They preserve $\mathcal{N}=1$ supersymmetry, and such manifolds are called weak $G_2$- or nearly-parallel $G_2$-manifolds in the mathematics literature. 
All smooth homogeneous weak $G_2$-manifolds have been classified \cite{Friedrich:1997iau}. Moreover, it has been found that some massless type IIA solutions compactified on twistor spaces, i.e. special cases of the coset spaces $\text{Sp(2)/Sp(1)$\times$U(1)}\cong \mathbb{CP}^3$ and $\text{SU(3)/U(1)$\times$U(1)}\cong \mathbb{F}(1,2,3)$ \cite{Tomasiello:2007eq,Aldazabal:2007sn,Koerber:2008rx,Caviezel:2008ik}, lift to M-theory compactified on weak $G_2$-spaces, i.e. the squashed seven-sphere $\text{Sp(2)/Sp(1)}\cong \tilde{S}^7$ and the Aloff-Wallach space $\text{SU(3)/U(1)}\cong N_{p,q}$ respectively (where $p$ and $q$ are integers parametrising the embedding of the U(1)). These type IIA compactifications do not require sources. However, it is interesting to see whether something similar can happen in the presence of D-brane or O-plane sources. 

Indeed, all this leads one to wonder whether the 7d manifold constructed in ref.~\cite{Cribiori:2021djm} exhibits weak $G_2$-holonomy as well. This is not immediately clear from the local type IIA solution that was constructed by solving the 10d equations of motion perturbatively, as done in ref.~\cite{Cribiori:2021djm}. The reason is that $\mathcal{N}=1$ supersymmetry cannot be retained automatically by this approach, and hence it is unclear whether a lift to a supersymmetric solution of M-theory appears. To overcome this issue, one could construct the local solution by solving the pure spinor equations perturbatively instead of the 10d equations of motion. This is precisely what this paper is achieving. It is shown that, at first order in the perturbation scheme, it can be obtained within an SU(3)-structure, which can then be lifted to a 7d $G_2$-structure in M-theory. 

The paper is structured as follows: in section \ref{sec:mlIIA_review}, the $\mathcal{N}=1$ AdS$_4$ solutions of massless type IIA are reviewed in the SU(3)-structure language. Section \ref{sec:WeakG2_Mtheory} discusses M-theory compactifications on 7d manifolds with $G_2$-structure and illustrates how the massless type IIA solutions with an SU(3)-structure manifold lift to such M-theory solutions.
The details of the localisation procedure with the pure spinor techniques of the scale-separated type IIA solution are discussed in section \ref{sec:O6_localisation}. The interested reader is also referred to appendix \ref{app:O6_localisation_one-forms} for more details on this approach. The vacuum then lifts to an M-theory solution on a manifold with weak $G_2$-structure which has the scale separation property.
Section \ref{sec:R_lambda_discussion} discusses some of the geometric aspects of this proposed manifold, and a summary can be found in section \ref{sec:summary}.

\section{Review of \texorpdfstring{AdS$_4$}{AdS_4} compactifications of massless type IIA on SU(3)-structure manifolds}
\label{sec:mlIIA_review}
This section gives a brief review of the scale-separated solution of type IIA supergravity compactified on the Iwasawa manifold. For this paper, it is useful to write the 10d metric in string frame as follows:
\begin{equation}
    \dd s_{10}^2 = \e^{2A}\dd s_4^2+\dd s_6^2\,,
\end{equation}
where $A$ is the warp factor. The Ramond-Ramond fluxes can be split into their external and internal components:
\begin{equation}
    \hat{F} = \e^{4A}\vol_4\wedge\star_{6} \lambda F + F,
\end{equation}
where the polyform $F$ is completely internal and $\lambda$ is the reversal operator. One can then solve the pure spinor equations on an SU(3)-structure background, which was mainly developed in refs.~\cite{Grimm:2004ua,Grimm:2018cpv,Lust:2004ig,Grana:2004bg,Behrndt:2004mj,Behrndt:2004km,Grana:2005sn}. The solutions are given as follows:
\begin{align}
\label{eq:H}
H_3 &= 0\,,\\
\label{eq:F0}
g_s F_0 &= 0\,,\\
\label{eq:F2}
g_s F_2 &=  -5 \tilde{m} \e^{-4A} J  - J^{-1}\llcorner \dd \left(\e^{-3A} \Im \Omega \right)\,,\\
\label{eq:F4}
g_s F_4 &= 0\,,\\
\label{eq:F6}
g_s F_6 &=  3\tilde{m}\e^{-4A} \vol_6\,.
\end{align}
It is important to note that $\e^{3A-\phi}=g_s^{-1}$ is constant. The parameter $\tilde{m}$ also fixes the 4d AdS radius $L_H^2 = \tilde{m}^{-2}$ and some of the torsion classes. The operator $J^{-1}\llcorner$ contracts the form that it is acting on with the elements of the inverse of $J$ and its definition and properties can be found in appendix \ref{app:generalities}. The only non-vanishing torsion classes are $\mathcal{W}_1$, $\mathcal{W}_2$ and $\mathcal{W}_5$, which parametrise the non-closure of $J$ and $\Omega$ as follows:
\begin{align}
	\label{eq:dJ}
 	\dd J &= \frac{3}{2}\Im\left(\overline{\mathcal{W}_1} \Omega\right),\\
	\label{eq:dOmega}
	\dd \Omega &= \mathcal{W}_1 J \wedge J + \mathcal{W}_2 \wedge J + \mathcal{W}_5 \wedge \Omega \,,
\end{align}
 and the torsion classes satisfy
\begin{equation}
\label{eq:torsion_solutions}
    \mathcal{W}_1 = -i \frac{4}{3} \: \e^{-A} \tilde{m}, \qquad \mathcal{W}_5 = \dd A,  \qquad  \Re \mathcal{W}_2 = 0\,.
\end{equation}
The curvature of the 6d manifold is completely determined by the torsion classes and can be written as (e.g. found in \cite{Larfors:2013zva}) 
\begin{equation}
    R_6 = \frac{15}{2}|\mathcal{W}_1|^2 -\frac{1}{2}|\mathcal{W}_2|^2  + 2 \star \dd \star \mathcal{W}_5\,.
\end{equation}
All of these equations have to be supplemented with the Bianchi identities for the RR fluxes, such that they all solve the 10d equations of motion. Only the Bianchi identity for $F_2$ is non-trivial. It is given by
\begin{equation}
\label{eq:F2_BI}
    \dd F_2 = \sum_i \delta_{\text{D6}_i} - 2 \sum_i \delta_{\text{O6}_i} \,,
\end{equation}
where the $\delta_{\text{D6/O6}_i}$ is the Dirac-delta three-form that localises the $i$-th D6/O6 source in its transverse space.
It is very important to remark that there is a priori no need for the warp factor and dilaton to be constant, unlike in massive type IIA where this was enforced by the Bianchi identity of $F_0$.
Nevertheless, in the absence of sources, or when they are treated in the smeared approximation, there is no need for non-trivial warping and the warp factor can be fixed to 1. In that case, the Bianchi identity takes the following form:
\begin{equation}
\label{eq:tadpole_specified}
   \frac{2 \tilde{m}^2}{3} \Re \Omega + i\dd \mathcal{W}_2  = \sum_i j_{\text{D6}_i}- 2 \sum_i j_{\text{O6}_i}\,.
\end{equation}
The quantity $j_{\text{D6/O6}_i}$ corresponds to the source three-form where the Dirac-delta source is smeared over the whole transverse space of the $i$-th D6/O6 source.

There are three interesting ways to solve the Bianchi identity. First, one can consider half-flat manifolds for which no sources are needed. This requires that the manifold has a $\mathcal{W}_2$-torsion class such that the left-hand side of eq.~\eqref{eq:tadpole_specified} vanishes.
Such scenarios have been studied for twistor spaces or specific coset spaces Sp(2)/((Sp(2)$\times$U(1)) and SU(3)/U(1)$\times$U(1), which are topologically $\mathbb{CP}^3$ and $\mathbb{F}(1,2,3)$ respectively \cite{Aldazabal:2007sn,Tomasiello:2007eq,Caviezel:2008ik,Koerber:2008rx}.

Second, one can consider nearly-K\"ahler manifolds (sometimes also referred to as manifolds with weak SU(3) holonomy) with wrapped D6-branes. Such manifolds have a vanishing $\mathcal{W}_2$-torsion class, hence requiring an overall positive contribution from the source. D6-branes suffice for that cause, although additional orientifold planes are not forbidden. 
All 6d homogeneous manifolds that admit a nearly-K\"ahler metric have been classified and the list is not extensive, it only comprises $S^6$, $S^3 \times S^3$, $\mathbb{CP}^3$ and $\mathbb{F}(1,2,3)$, the latter being the flag manifold (see e.g. \cite{Koerber:2010bx,Tomasiello:2022dwe} amongst the many references on this topic). Non-homogeneous nearly-K\"ahler manifolds also exist but have not been completely classified yet.
With appropriate choices of the fluxes, both weakly and strongly coupled solutions can be identified.

At last, one can consider cases where only O6-planes are necessary to solve the tadpole constraints. It is this type of SU(3)-structure background that is of most interest in this paper, as they allow for scale-separated solutions. The model that is studied in more detail here is the compactification on the 6d Iwasawa nilmanifold $\mathcal{M}_\text{Iw}$, which is an orbifold of a twisted six-torus, $\mathbb{T}^6/(\mathbb{Z}_2 \times \mathbb{Z}_2)$. Compactifications on this particular space were discussed earlier in refs. \cite{Banks:2006hg,Caviezel:2008ik,Koerber:2008rx} and analysed into more detail in ref.~\cite{Cribiori:2021djm}. Recently, ref.~\cite{Carrasco:2023hta} generalised this construction to similar compactifications on half-flat manifolds with similar properties. 
Compactifications using these spaces originate from their T-dual counterparts in massive type IIA string theory, where the manifolds are Calabi-Yau.

The compactification on the Iwasawa manifold is reviewed below, as it plays a central role in this paper.
The original solution uses smeared O6-planes such that the warp factor and dilaton are constant, i.e. $ \e^A =1$ and $\e^\phi =g_s$.
The SU(3)-structure forms $J$, $\Omega$ and the torsion class $\mathcal{W}_2$ of the Iwasawa manifold can be written as follows: 
\begin{align}
\label{eq:J-Omega-W2_Iw}
	J = J_1 + J_2 + J_3\,, \qquad
	\Omega =  \sum_{\alpha=1}^4 \left(k_{\perp,\alpha} + i k_{||,\alpha}\right) \,, \qquad
	 \mathcal{W}_2 = \frac{8 \tilde{m}}{3} i \left(2 J_1 - J_2 -J_3\right)\,,
\end{align}
where three-forms $k_{\perp, \alpha}$ and $k_{||, \alpha}$ satisfy $\star_6 k_{\perp, \alpha} = k_{||, \alpha} $, and the $k_{||,\alpha}$ are also the volume forms of the cycles that the O6-planes are wrapping, and the $k_{\perp,\alpha}$ are the volume forms of the cycles transverse to those. So there are four sets of O6-planes along different directions, each set comprising eight O6-plane images.
In terms of a basis of one-forms $e^i$, one can write the forms above as 
\begin{equation}
\label{eq:J_basis}
    J_1 =-L_T^{2} e^{16}, \qquad J_2 = - L_2^2e^{24}, \qquad J_3 = L_3^2 e^{35}\,,
\end{equation}
and the three-forms are 
\begin{align}
\label{eq:k_parallel}
        \left(k_{||,1}, k_{||,2}, k_{||,3} , k_{||,4}\right) &= L_T L_2 L_3 \; \left( e^{123},e^{145}, e^{256}, e^{346} \right)\,, \\
\label{eq:k_perp}
        \left(k_{\perp,1}, k_{\perp,2}, k_{\perp,3} , k_{\perp,4} \right) &= L_T L_2 L_3 \; \left( e^{456},e^{236}, -e^{134}, -e^{125}\right)\,.
\end{align}
The torsion classes are realised by the non-closure of the following one-forms,
\begin{equation}
\label{eq:geomflux}
    \dd e^{1} = -e^{23} - e^{45}\,,\qquad \dd e^{6} = - e^{34} - e^{25},
\end{equation}
and where all others are closed. The metric is written as 
\begin{equation}
    \dd s_6^2 = L_T^2 \left[ (e^1)^2 + (e^6)^2 \right]+L_2^2 \left[ (e^2)^2 + (e^4)^2 \right]+L_3^2 \left[ (e^3)^2 + (e^5)^2 \right]\,.
\end{equation}
Furthermore, the length scales are related by the pure spinor equations, such that
\begin{equation}
\label{eq:mtilde}
    2\tilde{m} = \frac{L_T}{L_2 L_3}\,.
\end{equation}
The solutions \eqref{eq:H}-\eqref{eq:dOmega} of the pure spinor equations profit from a scaling symmetry, realised by the following scalings:
\begin{align}
\label{eq:general_scalings}
\begin{split}
    &L_T \sim n^{(a-b-c)/4}, \qquad L_2 \sim n^{(a-b+c)/4}, \qquad L_3 \sim n^{(a+b-c)/4}\,,\\
    & L_H \sim n^{(a+b+c)/4}, \qquad g_s \sim n^{(a -3b - 3c)/4},
\end{split}
\end{align}
where all the length scales are expressed in string units and $c \geq b>0$ is taken without loss of generality. All $a,b,c$ need to be positive, as they also determine the unbounded fluxes. There are two $F_2$-fluxes and one $F_6$-flux that scale as follows:
\begin{equation}
    F_{2,2} \sim n^c, \qquad F_{2,3} \sim n^b, \qquad F_6 \sim n^{a}\,,
\end{equation}
where the two-form fluxes $F_{2,l}$ are proportional to $J_l$.
To determine whether scale separation is achieved, one has to compare the AdS radius $L_\text{H}=\tilde{m}^{-1}$ with the KK radius $L_\text{KK}$, which can be estimated by the largest radius of the manifold, $L_{2}$ in this case (see also ref.~\cite{Cribiori:2021djm} where the first eigenvalue of the Laplacian was estimated). The setup is indeed scale-separated, as
\be
	\frac{L_\text{KK}^2}{L_H^2} \sim n^{-b}.
\ee
Additionally, one needs to verify that the O6-backreaction is under control in the large $n$-limit. The strength of the O6-plane backreaction is measured by $g_s/\bar L$ \cite{Baines:2020dmu,Cribiori:2021djm}, where $\bar{L}$ is the backreaction radius, which in this setup is given by $L_T L_3/L_2$ \cite{Cribiori:2021djm}. Hence, the backreaction is suppressed in the large flux regime by:
\be
\label{eq:expansion_parameter}
	\frac{g_s}{\bar{L}} \sim n^{-b}.
\ee
Furthermore, to obtain a properly behaved solution where all cycles grow in the large flux regime, one has to require that 
\begin{equation}
    a > b+c \geq 2b.
\end{equation}
However, there is still enough freedom in choosing the scaling exponents, allowing both for weakly and strongly coupled solutions. Indeed, one has strong coupling when
\begin{equation}
    a > 3(b+c) \geq 6 b\,.
\end{equation}
Otherwise, one has weak coupling when 
\begin{equation}
    2b \leq b + c < a < 3(b+c)\,.
\end{equation}
The strongly coupled solutions are of interest here, as such setups can in principle be lifted to M-theory. This is discussed in the next section. 

\section{Weak \texorpdfstring{$G_2$}{G_2} holonomy and M-theory compactifications}
\label{sec:WeakG2_Mtheory}
In the previous section, 4d $\mathcal{N}=1$ compactifications have been discussed in type IIA. These arise from 6d manifolds with an SU(3)-structure as such manifolds allow for one Killing spinor. Likewise, for 7d manifolds, when there is only one Killing spinor, the manifold has a $G_2$-structure, which plays an important role in 4d $\mathcal{N}=1$ compactifications of M-theory. All the metric information of a $G_2$-structure manifold is encoded in the left-invariant three-form $\Phi$ and associate four-form $\star \Phi$. They do not need to be closed, which is described by the torsion classes $W_1$, $W_7$, $W_{14}$ and $W_{27}$ as follows:
\begin{align}
    &\dd \Phi = W_1 \star \Phi + W_7 \wedge \Phi + W_{27}\,,\\
    &\dd \star \Phi = \frac{4}{3}W_7 \wedge \star \Phi + W_{14} \wedge \Phi\,.
\end{align}
The index on the torsion classes indicates to which irreducible presentation of $G_2$ they belong.
The seven-dimensional curvature can also be expressed in terms of the torsion classes as well:
\begin{equation}
\label{eq:R7_general}
    R_7 = \frac{21}{8}|W_1|^2 + \frac{10}{3}|W_7|^2 + 4 \star \dd \star W_7 -\frac{1}{2} |W_{14}|^2 -\frac{1}{2} |W_{27}|^2 \,.
\end{equation}
When all the torsion classes vanish, the manifold is Ricci-flat and has $G_2$-holonomy. Non-compact spaces of this type can easily be constructed as cones with 6d nearly-K\"ahler manifolds as the base. 
Amongst compact examples are toroidal manifolds, which are classified by Joyce \cite{Joyce:1996i,Joyce:1996ii}.
Additionally, it was shown in ref.~\cite{Andriolo:2018yrz} that one can also construct compact spaces with $G_2$-holonomy from circle fibrations over 6d symplectic manifolds where only the $\mathcal{W}_2$- and $\mathcal{W}_5$-torsion classes could be non-vanishing. These were obtained by lifting 4d Minkowski solutions of type IIA to M-theory. Such solutions are also described by the equations \eqref{eq:H}-\eqref{eq:torsion_solutions}, but with $\tilde{m} = 0$. This also suggests that such manifolds with local D6- and/or O6-sources do lift to spaces with $G_2$-holonomy, albeit some potential singularities at the loci of the sources in the type IIA setting.

In this paper, manifolds with weak $G_2$-holonomy are of interest. For those, all torsion classes except $W_1$ vanish, and as a consequence, $W_1$ is also constant. They have a positive Einstein metric and hence allow for M-theory compactifications of the Freund-Rubin type. Compactifications on manifolds with weak $G_2$-holonomy were previously discussed in \cite{Behrndt:2003uq,Behrndt:2003zg,Lukas:2004ip,Acharya:2003ii}. They also preserve four supercharges, leading to $\mathcal{N}=1$ supersymmetry. All such smooth homogeneous spaces have been classified, being the squashed $\tilde{S}^7$, the Aloff-Wallach spaces $N_{p,q}$, and the Berger space SO(5)/SO(3) \cite{Friedrich:1997iau}.

Also interesting is that ref.~\cite{Bilal:2003bf} showed that one can construct compact manifolds with weak $G_2$-holonomy and two conical singularities by the use of nearly-K\"ahler manifolds.
Such constructions were also considered in ref.~\cite{Acharya:2003ii} for Freund-Rubin compactifications of M-theory. In that paper, such setups were mainly studied from the near-horizon limit of M2-brane backgrounds and their type IIA duals.

It is also known that some AdS$_4$ compactifications of $\mathcal{N}=1$ type IIA on SU(3)-structure manifolds lift to M-theory compactifications on manifolds with weak $G_2$-holonomy. These are the solutions on the half-flat manifolds that do not require local sources discussed in the previous section.

However, inspired by ref.~\cite{Andriolo:2018yrz}, compactifications with D6- and/or O6-sources also seem to lift to weak $G_2$-manifolds, albeit potential singularities.
To start, remember that one can lift a strongly coupled solution of massless type IIA to M-theory by fibering the 6d manifold over a circle. The 11d metric has the following form:
\begin{equation}
    \dd s_{11}^2 = \e^{-\frac{2}{3}\phi} \dd s_{10}^2 + \e^{\frac{4}{3}\phi} v^2\,,
\end{equation}
where $v = \dd z - \mathcal{C}$ is a one-form that encodes the additional circle parametrised by $z$ and the non-trivial connection $\mathcal{C}$, which is the local gauge potential of $F_2$. Hence, $v$ must satisfy
\begin{equation}
    \dd v = - F_2\,.
\end{equation}
A nice property of the type IIA solutions discussed above, is that the warping and dilaton profiles are related in such a way ($\e^{3A-\phi} = g_s^{-1}$) that the 4d metric is not warped anymore:
\begin{equation}
    \dd s_{11}^2 = g_s^{-\frac{2}{3}} \dd s_4^2 + \dd s_7^2, \hspace{2cm} \dd s_7^2 =  \e^{-\frac{2}{3}\phi}\dd s_{6}^2 + \e^{\frac{4}{3}\phi}v^2.
\end{equation}
The only other non-vanishing  RR flux $F_6$ lifts to the internal space-filling flux $G_7$ of M-theory as 
\begin{equation}
    G_7 = v \wedge F_6 \,.
\end{equation}
Furthermore, it is known that compactifications on SU(3)-structure manifolds lift to manifolds with $G_2$-structure. The $G_2$-invariant three- and four-forms in 7d can be expressed in terms of the 6d holomorphic three-form and K\"ahler two-form as follows:
\begin{align}
    \Phi &= - v \wedge J + \e^{-\phi}\; \Im \Omega\,,\\
    \star \Phi &= \e^{-\frac{4}{3}\phi} \frac{1}{2}J \wedge J - v \wedge \e^{-\frac{1}{3}\phi} \;\Re \Omega\,.
\end{align}
To preserve some conformity with the literature, it is more convenient to trade the dilaton for the warp factor, such that one can write
\begin{align}
\label{eq:Phi_def}
    \Phi &= - v \wedge J + g_s^{-1}\e^{-3A}\; \Im \Omega\,,\\
\label{eq:starPhi_def}
    \star \Phi &= g_s^{-\frac{1}{3}}\left(g_s^{-1}\e^{-4A} \frac{1}{2}J \wedge J - v \wedge \e^{-A} \Re \Omega\right)\,.
\end{align}
The goal is to show now that this $G_2$-structure has weak $G_2$-holonomy, meaning that $W_{7}$, $W_{14}$ and $W_{27}$ are vanishing but $W_1$ is not. 
In fact, $W_1$ is computed to be $W_1 = -2 \tilde{m} g_s^{1/3}$, meaning that
\begin{equation}
\label{eq:weakG2_general}
    \dd \Phi = -2 \tilde{m} g_s^{\frac{1}{3}} \star \Phi, \qquad \dd \star \Phi = 0.
\end{equation}
To obtain this, one could use the results of ref. \cite{Kaste:2003dh}, where the torsion classes of a 6d SU(3)-structure manifold in the presence of a non-trivial RR two-form $F_2$ were related to those of $G_2$-structure manifold in full generality. However, it is easier for the purposes of this paper to derive the result when directly applied to the type IIA background under consideration.
The first equation of \eqref{eq:weakG2_general} is satisfied only when
\begin{align}
\label{eq:WeakG2_condition1}
    &\dd J = 2\tilde{m} \;\e^{-A} \Re \Omega\,,\\
\label{eq:WeakG2_condition2}
     & g_s F_2 \wedge J + \dd \left(\e^{-3A}\Im \Omega\right) = -2 \tilde{m}\left(\frac{1}{2}\e^{-4A} J \wedge J\right)\,.
\end{align}
The first condition \eqref{eq:WeakG2_condition1} is automatically fulfilled for the background under consideration due to eq.~\eqref{eq:dJ}, whereas the second also directly follows from the expression of $F_2$ \eqref{eq:F2}. Indeed, by using the commutation relation \eqref{eq:J_commutator} for $J^{-1}$ and $J\wedge$, one notices that
\begin{align}
\notag
   g_s F_2 \wedge J &= J \wedge \left( - J^{-1}\llcorner \dd \left( \e^{-3A} \Im \Omega\right) - 5 \tilde m \e^{-4A} J\right)\\
\label{eq:F2wJ_temp}
    &= -\dd \left( \e^{-3A} \Im \Omega\right) - J^{-1} \llcorner \left(  J \wedge \dd \left( \e^{-3A} \Im \Omega\right)\right) - 5 \tilde{m} \e^{-4A} J\wedge J\,.
\end{align}
The second term can be rewritten as follows:
\begin{align}
\notag
    J \wedge \dd \left( \e^{-3A} \Im \Omega\right) &= \dd \left( J \wedge \e^{-3A} \Im \Omega\right)- \dd J \wedge \left( \e^{-3A} \Im \Omega\right)\\
\notag
    &= - 2 \tilde{m} \e^{-A} \Re \Omega \wedge \e^{-3A}\Im \Omega\\
    &= -2 \tilde{m} \e^{-4A} (-4 \vol_6) = -2 \tilde{m} \e^{-4A} \frac{2}{3}J^3\,,
\end{align}
and with $J^{-1} \llcorner J^3 = 3 J \wedge J$, one finds that
\begin{equation}
    - J^{-1} \llcorner \left(  J \wedge \dd \left( \e^{-3A} \Im \Omega\right)\right) = 4 \tilde{m} \e^{-4A}J\wedge J\,.
\end{equation}
When this is plugged into eq.~\eqref{eq:F2wJ_temp}, eq.~\eqref{eq:WeakG2_condition2} is indeed recovered.
Then, $\dd \star \Phi =0$ should follow automatically by the nilpotency of the exterior derivative, since $W_1$ is constant. Indeed, by acting with the exterior derivative on eq.~\eqref{eq:starPhi_def}, one finds
\begin{align}
    \dd \star \Phi &= g_s^{-\frac{1}{3}}\left(g_s^{-1}\frac{1}{2} \dd \left(\e^{-4A} J \wedge J \right) + F_2\wedge \e^{-A}\Re \Omega\right)\\
    &= -\frac{g_s^{-\frac{1}{3}}}{2\tilde{m}}\left(\dd \left( F_2 \wedge J\right) - F_2\wedge \dd J \right)\\
    &=-\frac{g_s^{-\frac{1}{3}}}{2\tilde{m}} \dd F_2 \wedge J\,.
\end{align}
In the second line, the exterior derivative on eq.~\eqref{eq:WeakG2_condition2} was used. The final result vanishes away from sources by the Bianchi identity of $F_2$ \eqref{eq:F2_BI}.
One concludes that the massless type IIA AdS$_4$ backgrounds with SU(3)-structure lift to Freund-Rubin compactifications of M-theory on 7d manifolds with weak $G_2$-holonomy, away from sources where
\begin{equation}
    \dd \Phi = -2 \tilde{m}g_s^{\frac{1}{3}} \star \Phi, \qquad \dd \star \Phi = 0.
\end{equation}
Note that the presence of a non-trivial warp factor ($\dd A = \mathcal{W}_5\neq 0$) and the torsion class $\mathcal{W}_2$ do not play any role here.

The type IIA compactifications that do not need sources were already discussed above. Indeed, as described in refs.~\cite{Tomasiello:2007eq,Aldazabal:2007sn,Koerber:2008rx,Caviezel:2008ik}, the models on the cosets $\text{Sp(2)/Sp(1)$\times$U(1)}\cong \mathbb{CP}^3$ and $\text{SU(3)/U(1)$\times$U(1)}\cong \mathbb{F}(1,2,3)$ do not need sources and those lift to squashed seven-spheres $\text{Sp(2)/Sp(1)}\cong \tilde{S}^7$ and Aloff-Wallach spaces $\text{SU(3)/U(1)}\cong N_{p,q}$ respectively.

When sources are needed to solve the tadpole, they must be genuinely local, simply smearing them does not suffice. The reason is that for smeared sources, $\dd F_2$ is non-vanishing everywhere, implying that $\dd^2 v$ is non-vanishing everywhere as well, which is inconsistent.
For nearly-K\"ahler manifolds where only D6-branes are wrapped to solve the tadpole, some of such M-theory lifts were already discussed. For example, in ref.~\cite{Acharya:2003ii}, the compactification $\text{AdS}_4 \times S^3 \times S^3$ was considered for the lift to M-theory with D6-branes wrapping three-cycles to cancel the tadpole, which was later investigated in ref.~\cite{Aldazabal:2007sn} in the smeared limit. Although ref.~\cite{Acharya:2003ii} alluded to a local treatment of the D6-branes by considering the flat space limit close to the branes, the details of the full backreacted solution on the compact space $S^3\times S^3$ have not been worked out yet. 

Finally, in this work, the example of most interest here is the scale-separated model in massless type IIA, described in ref.~\cite{Cribiori:2021djm} and generalised in ref.~\cite{Carrasco:2023hta}.
The former reference used the Iwasawa nilmanifold, and a strongly coupled regime with scale separation was identified there, as reviewed in the previous section. However, the setup was initially treated in the smeared approximation. As explained above, such solutions cannot be lifted to M-theory as $F_2$ is not closed locally (up to the isolated O6 singularities), and hence an 11d metric cannot be constructed. This can also be concluded from looking at the 7d Ricci curvature in terms of the 6d curvature and the RR two-form $F_2$, given for constant dilaton and warping by $R_7 = g_s^{2/3}(R_6 -g_s^{2}|F_2|^2/2)$. The seven-dimensional manifold must be positively curved as there are no sources in the M-theory setup \cite{Gautason:2015tig}, whereas the nilmanifold is negatively curved, leading to a contradiction.
Fortunately, it turns out that a local treatment of the O6-planes in the type IIA supergravity theory allows one to overcome these issues. 
In ref.~\cite{Cribiori:2021djm}, the O6-plane backreaction was taken into account, essentially by regarding the backreaction as small corrections to the expressions of the smeared approximation. 
This was done by writing all the fields as expansions in $1/n^b$ (see the discussion around eq.~\eqref{eq:expansion_parameter}), where the zeroth-order terms contribute to the smeared results, and the higher order terms encode the local backreaction. The equations of motion were solved up to first order, where the solution is determined by a linear superposition of Green's functions of the scalar Laplacian operator. Those generate the Dirac-delta distributions that localise each of the O6-planes separately. Eventually, this local solution was lifted to M-theory at the level of the metric and the magnetic seven-form flux. By using the 11-dimensional equations of motion, it was indeed verified that the seven-dimensional curvature was positive and that the whole solution is of the Freund-Rubin type. 
However, the precise geometry of the seven-dimensional manifold was not explored further, and neither was it shown that the local solution preserves supersymmetry.
Nevertheless, the analysis in this section suggests that this seven-dimensional manifold could have weak $G_2$-holonomy, provided that the local solution can be described in the SU(3)-structure language.
This was not done in ref.~\cite{Cribiori:2021djm}, but is achieved in the next section.

Finally, consider the Ricci curvature of the proposed 7d manifold. Using eq.~\eqref{eq:R7_general}, it becomes
\begin{equation}
\label{eq:R7_computed}
    R_7 = \frac{21}{8}|W_1|^2 = \frac{21}{2}\tilde{m}^2 g_s^{\frac{2}{3}}.
\end{equation}
As expected, the curvature is positive. All of these considerations suggest that in principle, a 6d manifold with negative Ricci curvature can be fibered over a circle to a manifold with positive curvature. The reason for this is that the torsion class $\mathcal{W}_2$ provides a negative contribution to the 6d curvature and hence can lead to overall negative curvature, whereas the 7d manifold is always positively curved. The backreacted Iwasawa manifold may provide an example, as shown in ref.~\cite{Cribiori:2021djm}. Moreover, note that the 7d curvature calculated in that paper is the same as here, which serves as a nice consistency check.

\section{Localising O6-planes on the Iwasawa background in the pure spinor language}
\label{sec:O6_localisation}
As explained above, the local O6-backreaction corrections of the scale-separated AdS$_4 \times\mathcal{M}_\text{Iw}$ solution was originally computed in ref. \cite{Cribiori:2021djm}.
This was done by solving the Einstein equations perturbatively, which was accomplished previously in \cite{Junghans:2020acz} for the scale-separated DGKT vacua of \cite{DeWolfe:2005uu,Camara:2005dc}. Completely orthogonally and simultaneously, ref.~\cite{Marchesano:2020qvg} solved the pure spinor equations and Bianchi identities of the same background perturbatively. The latter approach has the advantage that supersymmetry is kept manifestly, whereas the former does not but has the advantage of being valid for the non-supersymmetric skew-whiffed DGKT vacua as well. 
The strategy of the pure spinor approach was to solve the Bianchi identity of $F_2$ first, and then compare the result with the $F_2$ solution from pure spinor equations for SU(3)$\times$SU(3)-structure backgrounds, found in ref.~\cite{Saracco:2012wc}. Doing so enables one to read off the necessary first-order corrections to the warping and three-form component of the pure spinors, from which other corrections could be determined as well. It is important to emphasise that the O6-backreaction generates non-trivial warping, which cannot arise in SU(3)-structure background with non-vanishing Roman's mass. For this reason, ref.~\cite{Marchesano:2020qvg} localised the DGKT vacua into SU(3)$\times$SU(3)-structure backgrounds.

The goal is to follow the same strategy for the scale-separated AdS$_4 \times$ Iwasawa solution.  The only difference is that at least to first order in the backreaction, one can stay within the SU(3)-structure Ansatz as non-trivial warping is allowed when the Roman's mass is vanishing.
The leading (zeroth) order solution to the system \eqref{eq:H}-\eqref{eq:torsion_solutions} is then given by the smeared approximation as presented in eqns.~\eqref{eq:J-Omega-W2_Iw}-\eqref{eq:geomflux}.

In ref.~\cite{Cribiori:2021djm}, the solution to the Bianchi identity of $F_2$ was already found in terms of the uncorrected SU(3)-structure data, to first order in the backreaction. At this order, the full solution including all O6-planes can be seen as a linear superposition of single sources.
From now on, the special scaling situation $c=b$ is chosen to simplify notation and such that $L_2 \sim L_3 \sim n^{a/4}$ and $L_T \sim n^{(a-2b)/4}$. The backreaction scale $\bar{L}$ is then $L_T$ in eq.~\eqref{eq:expansion_parameter}, up to an order-one constant $L_2/L_3$ which will be omitted for notational simplicity as it can also be absorbed in the quantities defined below.

Before presenting the result, one must introduce a couple of functions $\varphi_i$ and $\chi_i$, the former being related to the Green's function that generates the delta function on the space transverse to the $i$-th O6-plane, and the $\chi_i$ being related to $\varphi_i$ as follows
\begin{equation}
    \Delta \varphi_i = \frac{\tilde{m}}{L_T}(\delta_{\text{O6},i}-j)\,, \hspace{2cm}
    \Delta \chi_i = 8 \tilde m \frac{\varphi_i}{L_T} \,,
\end{equation}
where $j= 5L_T^2 \tilde{m}/(2g_s)$ is a constant that does not scale with $n$.
The sums of all these functions are denoted without an index,
\begin{equation}
    \varphi = \sum_i \varphi_i\,, \hspace{2cm} \chi = \sum_i \chi_i \, .
\end{equation}
Finally, one should introduce a three-form $k_\varphi$ such that
\begin{equation}
    \frac{\varphi}{L_T} \Re \Omega + \Re k_\varphi = \sum_i 4 \frac{\varphi_i}{L_T} k_{\perp,i} \,,
\end{equation}
where $k_{\perp,i} $ is the volume form of the cycle transverse to the $i$-th O6-plane, as defined in eq.~\eqref{eq:k_perp}.

It is very important to note that derivatives on the $\varphi_i$'s and $\chi_i$'s along the 1- and 6- directions, which have associate one-forms that are not-closed (cf. \eqref{eq:geomflux}), are neglected. The reason is that in the large $n$-limit, the associated length scale $L_T$ becomes much smaller than $L_2$ and $L_3$, such that derivatives of the Green's functions become exponentially suppressed in $n$, as was argued in ref.~\cite{Cribiori:2021djm}.
This means that only derivatives on the directions 2, 3, 4 and 5 are relevant.

Now that all necessary ingredients are introduced, one can finally write down the expression for the RR field strength $F_2$,
\begin{equation}
\label{eq:F2_expansion}
    F_2 = F_2^{(0)} + F_2^{(1)} + \cdots
\end{equation}
where $F_2^{(0)}$ is the expression from the smeared approximation. Similar notation is used for the SU(3)-structure data $J$ and $\Omega$:
\begin{align}
    J &= J^{(0)} + J^{(1)} + \cdots\\
    \Omega &= \Omega^{(0)} + \Omega^{(1)} + \cdots
\end{align}
where $J^{(0)}$ and $\Omega^{(0)}$ correspond to the K\"ahler two-form and holomorphic three-form on the Iwasawa manifold \eqref{eq:J-Omega-W2_Iw} respectively.
In terms of those, $F_2^{(0)}$ is just
\begin{equation}
\label{eq:F2(0)}
    F_2^{(0)} = -g_s^{-1}5\tilde{m} J^{(0)} -g_s^{-1}J^{(0)-1}\llcorner \dd \, \Im \Omega^{(0)}\,.
\end{equation}
The first order correction $F_2^{(1)}$ was already computed in ref.~\cite{Cribiori:2021djm} such that it solves the Bianchi identity and the 10d RR equation of motion. It was found to be
\begin{equation}
\label{eq:full_F2_corr}
    F_2^{(1)} = - J^{(0)-1} \llcorner \dd \left( 4 \frac{\varphi}{L_T} \Im \Omega^{(0)} - \Im k_\varphi \right) + 4\frac{\varphi}{L_T} \, J^{(0)-1} \llcorner (\dd \, \Im \Omega^{(0)}) + \frac{5}{2}\dd \mathcal{C}_1 \,,
\end{equation}
where the imaginary part of the form $k_\varphi$ is defined such that
\begin{equation}
\label{eq:k_chi_definition}
    4 \frac{\varphi}{L_T} \Im \Omega^{(0)} - \Im k_\varphi = \sum_i 4\frac{\varphi_i}{L_T}\left( \Im \Omega^{(0)} - k_{||,i} \right)\,.
\end{equation}
The one-form $\mathcal{C}_1$ is defined as follows
\begin{equation}
\label{eq:C1_def}
    \mathcal{C}_1  = -I^{(0)} \cdot \dd \chi\,,
\end{equation}
where $I^{(0)}$ corresponds to the almost complex structure of the Iwasawa manifold, see appendix~\ref{app:generalities} for its definition and properties.
Note that the exact term in eq.~\eqref{eq:full_F2_corr} was obtained by solving the 10d equation of motion of $F_2$, which was originally not derived in the pure spinor language in ref.~\cite{Cribiori:2021djm}.
Moreover, an important property of the exact form is that it satisfies
\begin{equation}
    J^{(0)-1}\llcorner \dd \mathcal{C}_1 = 8 \tilde{m} \frac{\varphi}{L_T}\,.
\end{equation}
Finally, the expression for $F_2$ above satisfies the Bianchi identity in the presence of O6-plane sources, i.e. 
\begin{equation}
    \dd F_2 = \dd F_2^{(0)}+ \dd F_2^{(1)} = -2\sum_i \delta_{\text{O6},i}\,.
\end{equation}
Now that the Bianchi identity is solved, one can investigate how this fits into the SU(3)-structure equation for $F_2$, i.e. eq.~\eqref{eq:F2}.

\subsubsection*{The corrections of the SU(3)-structure forms}

To proceed, the strategy is to find the backreaction corrections to the K\"ahler two-form $J$ and holomorphic three-form $\Omega$ and non-trivial warp factor $\e^A$, such that the SU(3)-structure identities are solved, at least to first order:
\begin{align}
\label{eq:ToBeSolved_1}
    &F_2^\text{SU(3)} - F_2^\text{BI} = \mathcal{O}\left(n^{-b}\right)\,,\\
\label{eq:ToBeSolved_2}
    &\dd J - 2 \tilde{m} \;\e^{-A}\Re \Omega = L_T^2 \; \mathcal{O}\left(n^{-2b}\right)\,.
\end{align}
Here $F_2^\text{BI}$ corresponds to the expressions \eqref{eq:F2_expansion}-\eqref{eq:full_F2_corr} that were obtained in ref.~\cite{Cribiori:2021djm} from solving the Bianchi identity and $F_2^\text{SU(3)}$ is the expression of the SU(3)-structure solution, eq.~\eqref{eq:F2}, which is repeated here for convenience:
\begin{equation}
\label{eq:F2SU3}
    F_2^\text{SU(3)} =  -5 \tilde{m} g_s^{-1} \e^{-4A} J  - g_s^{-1}J^{-1}\llcorner \dd \left(\e^{-3A} \Im \Omega \right)\,.
\end{equation} 
The reason for solving the equation at the indicated order, is that the terms in $F_2$ relevant for the Bianchi identity are at order $n^{0}$, whereas the ones for the non-closure of $J$ are at order $L_T^2 n^{-b}$.
At zeroth order, all the expressions must reduce to the ones in the smeared limit. Writing down the expansions for the SU(3)-structure forms is not as straightforward as in the Calabi-Yau case of ref.~\cite{Marchesano:2020qvg}, as the manifold is anisotropic. For instance, not all components of $J^{(0)}$ scale in the same way with $n$, whereas all components of $\Omega^{(0)}$ do. When focusing on $J^{(0)}$ in particular, one sees that it behaves as follows:
\begin{equation}
    J^{(0)} = n^{\frac{a}{2}}\left( n^{-b}\tilde{J}_1 + \tilde{J}_2 + \tilde{J}_3\right),
\end{equation}
where the tilded forms do not scale with $n$. However, only $J_1$ is responsible for $\dd J  = 2 \tilde{m} \Re \Omega$, which is subleading with respect to $J_2$ and $J_3$.
Hence, if one allows for first order corrections in $\Omega$ at $n^{-b}$, it means that one should include first-order corrections in $J$ that can look like second-order corrections on $J_2$ and $J_3$, i.e. that the first order correction $J^{(1)}$ could include the following terms
\begin{equation}
    J^{(1)} \supset n^{-b} \Phi_1 J_1 + n^{-2b} \Phi_2 J_2 + n^{-2b} \Phi_3 J_3\,,
\end{equation}
where the $\Phi_i$ parametrise the corrections, such that all terms in $J^{(1)}$ indeed scale the same as $n^{a/2-2b}$. For the three-form $\Omega$, such complications do not arise, although it turns out that one needs some second-order corrections as well, as is explained later in this section.

On top of the requirements \eqref{eq:ToBeSolved_1}-\eqref{eq:ToBeSolved_2}, the Monge-Ampère equation also must be satisfied:
\begin{equation}
\label{eq:Monge_Ampere}
    i \; \Omega \wedge \bar{\Omega} = \frac{4}{3}J\wedge J \wedge J = - 8 \vol_6 \,.
\end{equation}
It is easiest to account for the latter by realising that $J$ and $\Omega$ must be decomposable in terms of three complex one-forms, and find the corrections to those. In that sense, eq. \eqref{eq:Monge_Ampere} is trivially satisfied. This is precisely what is done in appendix \ref{app:O6_localisation_one-forms}, to which the interested reader is referred to for a more extensive discussion. In the remainder of this section, the results are given in terms of the warp factor $\e^{-4A}$ and the two- and three-forms $J$ and $\Omega$. The warp factor can be written as
\begin{equation}
    \e^{-4A} = 1 + 4 \frac{g_s \varphi}{L_T} + \cdots
\end{equation}
Additionally, the correction to the K\"ahler two-form $J$ is
\begin{align}
\begin{split}
    J^{(1)} &= \sum_i \left[g_s J^{(0)-1}\llcorner \left( \dd \chi_i \wedge( \Im \Omega^{(0)}- k_{||,i})\right) -  \frac{5}{8}\: g_s \chi_i J^{{(0)}-1}\llcorner \dd \, \Im \Omega^{(0)}\right]\\
    &= g_s J^{(0)-1}\llcorner \dd \left( \chi \; \Im \Omega^{(0)}- \Im k_\chi \right) -  \frac{11}{8}\: g_s \chi J^{{(0)}-1}\llcorner \dd \, \Im \Omega^{(0)}\,.
\end{split}
\end{align}
where $\Im k_\chi$ is defined as in \eqref{eq:k_chi_definition} but by replacing $4\varphi_i/L_T \to \chi_i$.
For the holomorphic three-form, it is convenient to include an overall warp factor in the corrections, such that
\begin{align}
\label{eq:OmegaI_cor}
\begin{split}
    (\e^{-3A}\Im\Omega)^{(1)}+(\e^{-3A}\Im\Omega)^{(2)} &= \sum_i\left[\frac{4g_s\varphi_i}{L_T} ( \Im \Omega^{(0)}- k_{||,i}) -  I^{(0)} \cdot \left(\frac{g_s }{2 \tilde{m}}\dd \chi_i \wedge F_2^{(0)} \right)\right]\\
    &= g_s \left(4 \frac{\varphi}{L_T}\Im \Omega^{(0)} - \Im k_\varphi \right) - I^{(0)}\cdot \left(\frac{g_s}{2\tilde{m}}\dd \chi \wedge F_2^{(0)} \right)\,,
\end{split}\\
\begin{split}
    (\e^{-A}\Re \Omega)^{(1)}+(\e^{-A}\Re \Omega)^{(2)} &= \sum_i \left[\frac{4 g_s \varphi_i}{L_T} k_{\perp,i} + \frac{g_s}{2 \tilde m}\dd \chi_i \wedge F_{2}^{(0)} \right]\\
    &= g_s \left(\frac{\varphi}{L_T}\Re \Omega^{(0)} + \Re k_\varphi \right) + \frac{g_s}{2\tilde{m}}\dd \chi \wedge F_2^{(0)}\,.
\end{split}
\end{align}
Note that the corrections to $\Omega$ involving the $\varphi_i$'s are the same as in the massive type IIA case with Calabi-Yau geometries of ref.~\cite{Marchesano:2020qvg}. Additionally, there are new corrections that did not appear there, i.e. the ones containing the $\chi_i$'s. These corrections appear as being wedged with $F_2^{(0)}.$
Note that the expressions of $\Im \Omega$ (and consequentially $\Re \Omega$) also contains second order contributions, i.e. at order $L_T L_2 L_3 n^{-2b}$. These come from the component of $F_2^{(0)}$ that does not scale, i.e. $F_{2,1}^{(0)} = -5 \tilde{m} g_s^{-1} J_1$.
This is quite surprising, as second-order contributions typically would result in second-order contributions in $F_2^\text{SU(3)}$ (i.e. at $\mathcal{O}(n^{-b})$) at which the equations are not solved. But this is not true for all such corrections. The key to this is that in eq.~\eqref{eq:F2SU3}, the operator $J^{-1}\llcorner$ contracts $\d (\e^{-3A} \Im \Omega)$ with the inverse elements of $J$. Because $J$ has components that scale differently, and as the $J^{-1}\llcorner$ operator inverts the length scales, it can enhance the scaling of some components of the form it is acting on. This is precisely the case here and is illustrated in more detail in appendix~\ref{app:O6_localisation_one-forms}.

To illustrate how the corrections (especially eq.~\eqref{eq:OmegaI_cor}) solve eq.~\eqref{eq:ToBeSolved_2}, consider writing
\begin{equation}
   \frac{1}{2 \tilde m} I^{(0)} \cdot \left( \dd \chi \wedge F_2^{(0)} \right) =\frac{1}{2 \tilde m} \left( I^{(0)} \cdot \dd \chi\right) \wedge F_2^{(0)}  = - \frac{1}{2\tilde{m}} \mathcal{C}_1 \wedge F_2^{(0)} = \Xi_1 + \Xi_2 \,,
\end{equation}
where the identity $I^{(0)} \cdot F_2^{(0)} = 0$ and eq.~\eqref{eq:C1_def} were used, and the three-forms $\Xi_1$ and $\Xi_2$ were defined as follows:
\begin{align}
    &\Xi_1 = -\frac{5}{2}\mathcal{C}_1 \wedge J^{(0)}\,, \\
    &\Xi_2 = -\frac{1}{2 \tilde m}\mathcal{C}_1 \wedge \left(J^{(0)-1} \llcorner\dd(\Im \Omega^{(0)})\right)\,.
\end{align}
Then using eqns.~\eqref{eq:C1_def} and \eqref{eq:IJv_props}, one notices that 
\begin{equation}
    \dd \Xi_1 = \frac{5}{2}\left[-\dd \mathcal{C}_1 \wedge J^{(0)} + 2 \tilde{m} \left(I^{(0)} \cdot \dd \chi \right) \wedge \Re \Omega \right]  =-\frac{5}{2}\left[\dd \mathcal{C}_1 \wedge J^{(0)} + 2 \tilde{m} \dd \chi \wedge \Im \Omega^{(0)}\right]\,.
\end{equation}
Then contracting the first term with $J^{(0)-1}$ gives
\begin{equation}
    J^{(0)-1}\llcorner \left[J^{(0)} \wedge \dd \mathcal{C}_1 \right] = \dd \mathcal{C}_1 + J^{(0)} \wedge  \left(J^{(0)-1}\llcorner \dd \mathcal{C}_1\right) = \dd \mathcal{C}_1 + 8\tilde{m} \frac{\varphi}{L_T} J^{(0)}. 
\end{equation}
For $\Xi_2$, it is best to realise that $J^{(0)-1} \llcorner\dd(\Im \Omega^{(0)}) = - 8 \tilde{m} (J_2 + J_3)$ and that the exact form $\dd \mathcal{C}_1$ also has no legs in common with $J_1$. This means that $\dd \mathcal{C}_1$ lives effectively on a four-dimensional subspace, such that in the equation, $J^{(0)-1}\llcorner$ can be replaced with $(J_2+J_3)^{-1}$ without any issue. Additionally, $(J_2+J_3)$ then defines a K\"ahler two-form on a 4d subspace and it satisfies, analogous to eq.~\eqref{eq:J_commutator}, $ [(J_2+J_3)^{-1}\llcorner, (J_2+J_3) \wedge] \alpha_p = (2-p)\alpha_p$, for any $p$-form $\alpha_p$ living in that 4d subspace. In that way, one computes that
\begin{align}
    J^{(0)-1}\dd \Xi_2 = - \frac{1}{2 \tilde{m}} (J_2+J_3)^{-1}\llcorner\left( \dd \mathcal{C}_1 \wedge - 8 \tilde{m} (J_2 + J_3)\right) =  -4 \frac{\varphi}{L_T} J^{(0)-1} \llcorner\dd \, \Im \Omega^{(0)}\,.
\end{align}
Combining all these equations and plugging them back into $F_2^\text{SU(3)}$ \eqref{eq:F2SU3}, the final result becomes indeed
\begin{equation}
    F_2^\text{SU(3)} = - J^{(0)-1} \llcorner \dd \left( 4\frac{\varphi}{L_T} \Im \Omega^{(0)} - \Im k_\varphi \right) + 4\frac{\varphi}{L_T} \, J^{(0)-1} \llcorner ( \dd \, \Im \Omega^{(0)} ) + \frac{5}{2}\dd \mathcal{C}_1 + \mathcal{O}(n^{-b})\,,
\end{equation}
where $\mathcal{O}(n^{-b})$ contains the remaining terms that contain single or no derivatives on $\chi$. In appendix \ref{app:O6_localisation_one-forms}, it is also verified that the metric resulting from the corrected SU(3)-data, is the same as the metric found in ref.~\cite{Cribiori:2021djm} at first order. This serves as a nice consistency check of the results obtained here. Additionally, there are non-diagonal terms generated at a lower order, i.e. $\mathcal{O}(n^{-3b/2})$, which are crucial for solving the non-diagonal components of the 10d Einstein equations, which were not considered in ref.~\cite{Cribiori:2021djm}. More details can be found in appendix~\ref{app:O6_localisation_one-forms}. 

With all these observations, it is clear that at least to first order in the perturbation theory, the setup allows for a local description of the O6-backreaction in an SU(3)-structure. From the results of section \ref{sec:WeakG2_Mtheory}, it then follows that the solution lifts in M-theory to a Freund-Rubin compactification on a weak $G_2$-manifold, again at least to first order in the backreaction. However, one has to remain open to the idea that at higher orders, the SU(3)-structure Ansatz might not suffice and one needs a generalised SU(3)-structure in 6d, as was used in ref.~\cite{Saracco:2012wc}.

\section{Comments on the Ricci curvature, diameter and lowest eigenvalue of the Laplacian}
\label{sec:R_lambda_discussion}
Now that it is shown that the scale-separated vacuum of massless type IIA lifts into a Freund-Rubin compactification of M-theory, it is time to discuss some properties of the proposed seven-dimensional manifold with weak $G_2$-holonomy.
As was discussed in section \ref{sec:WeakG2_Mtheory}, the Ricci scalar of the 7d manifold was easily computed in \eqref{eq:R7_computed}. It was shown in ref.~\cite{Gautason:2015tla} that for M-theory setups without sources, the curvature must satisfy a certain bound which is fulfilled here indeed (using $R_4 = 4 (-3\tilde{m}^2)g_s^{2/3}$):  
\begin{equation}
\label{eq:curvature_ratios}
    \left|\frac{R_7}{R_4}\right| = \frac{7}{8} <\frac{5}{4}\,,
\end{equation}
Note that this ratio does not scale with $n$, meaning there is no separation of scales between the 7d and the 4d curvature, the latter being equivalent to the AdS scale. 
For many manifolds though, the curvature scale is parametrically the same as the Kaluza-Klein scale, e.g. for spheres. A good proxy for the Kaluza-Klein scale is typically the lowest eigenvalue $\lambda_1$ of the Laplacian acting on scalars, at least to determine its parametric behaviour. This is typically estimated as the volume scale in isotropic settings, but more generally, in anisotropic settings, one should use the volume scale of the largest cycle in the manifold. 
In ref.~\cite{Cribiori:2021djm}, the lowest eigenvalue of the 6d Iwasawa manifold was computed by solving the Laplacian equation, and was indeed determined by the largest length scale $\lambda_1 \sim 1/L_2^2$. Although not taking backreaction effects into account, it was then argued that for the 7d manifold, the same happens, and since the additional circle is not larger than the other cycles, the lowest eigenvalue is determined by $L_2$ again. This does not scale in the same way as the curvature radius, and hence the Ricci scalar and first eigenvalue decouple parametrically. 
A similar story happens for the 6d base manifold, which is a nilmanifold with negative curvature instead. Its curvature can be calculated easily and is
\begin{equation}
    R_6 = - 8 \tilde{m}^2 = -2 \left(\frac{L_T}{L_2 L_3}\right)^2 \,,
\end{equation}
which does not scale the same as $1/L_2^2$ either because of the anisotropy of the manifold. 

The fact that this seems to happen for the 7d manifold as well, allows for scale separation despite eq.~\eqref{eq:curvature_ratios}. Indeed, the scale separation condition for these M-theory setups becomes the simple geometric constraint 
\begin{equation}
\label{eq:R7_lambda1_bound}
    R_7 \ll \lambda_1\,,
\end{equation}
while the curvature is positive.
This is a property that, at least to the knowledge of the author, has not been observed in the literature before for positively curved manifolds, especially not for manifolds with weak $G_2$-holonomy. Nevertheless, it is not excluded either. However, mathematically rigorous bounds have been formulated, relating the Ricci curvature, the lowest eigenvalue of the Laplacian operator and the diameter $D$ of a compact space $\mathcal{M}_\mathfrak{n}$ of dimension $\mathfrak{n}$.  This is defined as the maximal (geodesic) distance between all two points on the manifold,
\begin{equation}
    D = \sup_{p_1, p_2}\left[d(p_1, p_2) | p_1, p_2 \in \mathcal{M}_\mathfrak{n} \right]\,.
\end{equation}
The diameter allows for an estimate of the lowest eigenvalue of the Laplacian operator, given by an upper bound on $\lambda_1 D^2$ by a result of Cheng \cite{Cheng:1975} and a lower bound by Li and Yau \cite{Li:1980aa}, the latter being improved by Wu \cite{Wu:1991} for manifolds with non-negative curvature. The bounds state that 
\begin{equation}
\label{eq:ChengWu}
   \pi^2 \leq \lambda_1 D^2 \leq 2\mathfrak{n}(\mathfrak{n}+4) \,.
\end{equation}
This means that the first eigenvalue of the Laplacian and the diameter of the manifold must behave parametrically the same. In other words, if one can calculate the diameter of a positively curved compact space, one knows what the parametric behaviour of the lowest eigenvalue of the Laplacian operator is.

On the other hand, for manifolds with a positive Einstein metric, the curvature satisfies the following inequality by Yang \cite{Yang:1999}:
\begin{equation}
    \frac{\pi^2}{D^2} + \frac{R_\mathfrak{n}}{4} \leq \lambda_1\,.
\end{equation}
This bound combines with the upper bound \eqref{eq:ChengWu} to
\begin{equation}
    R_\mathfrak{n} \leq  \frac{4(2\mathfrak{n}(\mathfrak{n}+4)-\pi^2)}{D^2}\,.
\end{equation}
This does not forbid the scale separation condition \eqref{eq:R7_lambda1_bound} either. In fact, for the 7d manifold to achieve this, it must satisfy
\begin{equation}
    R_7 \ll  \frac{4(154-\pi^2)}{D^2} \lesssim \, \frac{60 \pi^2}{D^2}.
\end{equation}
Hence the diameter must become parametrically smaller than the curvature radius, as stated before. In ref.~\cite{Collins:2022nux}, it was conjectured that manifolds with this property do not exist, essentially by suggesting the existence of a lower bound that cannot be made parametrically small compared to the curvature. It was shown that it holds for K\"ahler-Einstein manifolds and checked for an array of manifolds. However, they either are isotropic and/or all preserve more than four supercharges. 
It therefore seems that the construction of the 7d manifold in this paper could be a counterexample to the conjecture. It would be very interesting to verify this statement by computing the diameter of the weak $G_2$-manifold proposed in this paper, which is beyond the scope of this work.
Note that similar bounds appeared in ref.~\cite{DeLuca:2022wfq} and other bounds on the spin-two spectrum have been formulated in refs.~\cite{DeLuca:2021ojx,DeLuca:2021mcj}.

Finally, one would hope to see whether one can construct (anisotropic) manifolds with weak $G_2$-holonomy with the scale separation property \eqref{eq:R7_lambda1_bound} but without the necessity of a perturbative construction, and potentially without singularities that originate from the O6-plane backreaction. 
Its importance is paramount because when one can find such a manifold with full mathematical rigour, one has established an indisputable Freund-Rubin vacuum with scale separation in M-theory.

\section{Summary}
\label{sec:summary}
This work provided evidence for the existence of scale-separated vacua of M-theory of the Freund-Rubin type, preserving minimal supersymmetry while the internal manifold has weak $G_2$-holonomy. This was achieved by lifting the scale-separated solutions of massless type IIA of ref.~\cite{Cribiori:2021djm} to M-theory. Such lifts require a local treatment of the O6-planes (instead of the usual smeared approximation), which was already considered in ref.~\cite{Cribiori:2021djm} at the level of the 10d equations of motion. However, the novelty of this work lies in performing the localisation of the O6-planes using the supersymmetry variations, in the form of pure spinor equations, allowing to express the backreaction corrections in terms of the SU(3)-structure data $J$ and $\Omega$. The corrections were calculated by solving the pure spinor equations and Bianchi identities in a perturbative approach, at first order in the O6-backreaction parameter. At that order, the manifold still retains an SU(3)-structure and manifestly preserves minimal supersymmetry. 
The power of this approach, as was explained in section \ref{sec:WeakG2_Mtheory}, is that the SU(3)-structure solution can be lifted directly into a $G_2$-structure of M-theory, and the manifold was shown to have weak $G_2$-holonomy, away from the original O6-plane loci.
The O6-planes lift into the geometry such that the M-theory solution does not contain any local sources. This solution exhibits the scale separation property, meaning that the 7d manifold with weak $G_2$-holonomy has Ricci curvature that becomes parametrically smaller than the first eigenvalue of the Laplacian, going against the conjecture formulated in ref.~\cite{Collins:2022nux}. For that reason, it would be very interesting to estimate the diameter of the 7d space. Moreover, it would be worthwhile to apply the same localisation techniques for the scale-separated solutions of ref.~\cite{Carrasco:2023hta} in finding a wider class of such 7d manifolds.

Nevertheless, the proposed 7d manifold here exhibits singularities originating from the O6-plane loci in type IIA. It is expected though that these codimension-four singularities can be resolved into Atiyah-Hitchin geometries. This was shown for the lift of one O6-plane in 10d with flat space asymptotics \cite{Atiyah:1988jp,Seiberg:1996nz}.
One might worry that such a reasoning might not go through at the intersection of the O-planes. However,
as was argued for in ref.~\cite{Junghans:2023yue}, the O6-planes do not necessarily have to intersect. This was shown in the context of smooth Calabi-Yau spaces, considering O6-planes on the blow-up of specific toroidal orbifolds. Usually, the multiple intersecting orientifold planes are generated by the orbifold images, but this is not necessarily true after the blow-up procedure, such that there is only one O-plane. This suggests the possibility of a similar phenomenon occurring for the O-planes on the Iwasawa manifold.
Another interesting avenue is to investigate the holographic dual to the proposed M-theory solution. Given that it is a Freund-Rubin vacuum on a manifold with weak $G_2$-holonomy, one naively expects that the holographic dual would be the worldvolume gauge theory of a stack of M2-branes, located at the tip of an eight-dimensional cone with Spin(7)-holonomy and filling a three-dimensional Minkowski spacetime. The base of the cone is then the weak $G_2$-manifold and the near-horizon limit near the M2-branes reduces to the scale-separated M-theory vacuum. M-theory on such Spin(7)-cones have been studied before, see e.g. refs.~\cite{Gukov:2001hf,Cvetic:2001zx,Cvetic:2001ye,Cvetic:2001pga} for early work. The investigation of the holographic setup along the lines of refs.~\cite{Apers:2022tfm,Conlon:2020wmc,Conlon:2021cjk,Ning:2022zqx,Apers:2022vfp,Apers:2022zjx} and whether it exhibits a parametrically large gap in the operator spectrum, is left for future research.

Although the main focus of this paper has been on the scale-separated type IIA solution and its lift to M-theory, 
the same can be considered for any other AdS$_4$ solution of massless type IIA on an SU(3)-structure background, with the manifold being half-flat. The analysis of this paper suggests that they also lift to manifolds with weak $G_2$-holonomy. Some of the sourceless type IIA backgrounds, i.e. on the twistor spaces that are topologically $\mathbb{CP}^3$ and $\mathbb{F}(1,2,3)$, were already known to lift to Freund-Rubin compactifications on the weak $G_2$-manifolds $\tilde{S}^7$ and $N_{p,q}$ respectively \cite{Tomasiello:2007eq,Aldazabal:2007sn, Caviezel:2008ik,Koerber:2008rx,Acharya:2003ii}. However, it would be very interesting to investigate lifts of backgrounds with solely D6-brane sources, e.g. on nearly-K\"ahler manifolds, which was partially explored in ref.~\cite{Acharya:2003ii}.

Finally, finding similar constructions of the 7d manifolds with weak $G_2$-holonomy and the scale separation property, but without the need to revert to perturbative techniques, would be particularly significant. In that way, one would realise a fully controlled M-theory set-up with $\mathcal{N}=1$ supersymmetry and scale separation.

\section*{Acknowledgements}
I would like to thank Magdalena Larfors and especially Alessandro Tomasiello and Thomas Van Riet for interesting discussions. I am also grateful to the Erwin Schr\"odinger Institute and their scientific program ``The Landscape vs. the Swampland'' where part of this work was completed. I am financially supported by the Olle Engkvists Stifelese.

\appendix
\section{Useful identities}
\label{app:generalities}

\subsection*{Definition and properties of \texorpdfstring{$I\cdot$ and $J^{-1}\llcorner$}{I\cdot and J^{-1}\llcorner}}
The almost complex structure $I$ and almost symplectic structure $J$ are regularly used in the main text. They can both be expressed in terms of the K\"ahler two-form $J$.
More precisely, the operators $J^{-1}\llcorner$ and $I \cdot$ are defined as follows:
\begin{equation}
\label{eq:Jcontract_Idot}
    J^{-1}\llcorner  =-\frac{1}{2}(J^{-1})^{ab}\iota_{e^b}\iota_{e^a}\,, \qquad {} \qquad  I \cdot = J_a {}^b e^a \wedge \iota_{e^b} \,
\end{equation}
where $\iota_{e^a}$ contracts the form it is acting on with the one-form $e^a$.
These operators satisfy the following identities:
\begin{gather}
    \frac{1}{6}\mathrm{Tr}(I^2) = -1\,, \hspace{3cm}  I \cdot J = 0\,, \hspace{3cm}
    J^{-1}\llcorner \Omega = 0\,,\\
     I  \cdot \Omega = 3 i  \Omega\,, \hspace{3cm}  J^{-1}\llcorner J = 3\,.
\end{gather}
The last identity is a special case of the following commutation relation:
\begin{equation}
\label{eq:J_commutator}
    \left[J^{-1}\llcorner, J\wedge \right]\alpha_p = (3-p)\alpha_p,
\end{equation}
where $\alpha_p$ is any $p$-form. The 3 appears as the complex dimension of the manifold. Operators that satisfy such a commutation relation are called Lefschetz operators \cite{Tomasiello:2022dwe}.
Additionally, for a one-form $v$, one has the following useful identities:
\begin{equation}
\label{eq:IJv_props}
    I \cdot v = - v \llcorner J, \qquad (I \cdot v) \wedge \Omega = - i v \wedge \Omega\,,
\end{equation}
where $v\llcorner = v_b\; g^{ab}\;\iota_{e^b}$.
For the Iwasawa manifold with the choice of one-forms used here, the operators $I^{(0)}\cdot$ and $J^{(0)-1}\llcorner$ act as follows: 
\begin{align}
    \begin{split}
    J^{(0)-1}\llcorner &=-\frac{1}{2}((J^{(0)})^{-1})^{mn}\iota_{e^n}\iota_{e^m} =  -\frac{1}{L_T^2}\iota_{e^6}\iota_{e^1} -\frac{1}{L_2^2}\iota_{e^4}\iota_{e^2} +\frac{1}{L_3^2}\iota_{e^5}\iota_{e^3}\,,
    \end{split}\\
    \begin{split}
    I^{(0)} \cdot & = (J^{(0)})_m {}^n e^m \wedge \iota_{e^n}\\
    &= e^6 \wedge \iota_{e^1} - e^1 \wedge \iota_{e^6} + e^4 \wedge \iota_{e^2}-e^2 \wedge \iota_{e^4}+e^3 \wedge \iota_{e^5}-e^5 \wedge \iota_{e^3}\,.
    \end{split}
\end{align}
It is good to clarify that the basis two- and three-forms are defined in eqns.~\eqref{eq:J_basis}-\eqref{eq:k_perp} satisfy the following identities:
\begin{equation}
    \dd J_{2,3} = 0\,, \qquad \dd J_1 = 2 \tilde{m} \Re \Omega^{(0)}\,.
\end{equation}
Moreover, the torsion classes are such that
\begin{equation}
    \dd \, \Im \Omega^{(0)} =  -8 \tilde{m} J_2 \wedge J_3\,,
\end{equation}
but more generally, for all the $k_{||,\alpha}$ and $k_{\perp,\alpha}$,
\begin{equation}
    \dd k_{||,\alpha} = -2 \tilde{m} J_2 \wedge J_3\,, \qquad \dd k_{\perp,\alpha} = 0\,.
\end{equation}
With this, one also notices that 
\begin{equation}  
    J_2 + J_3 = -\frac{1}{8\tilde{m}}J^{(0)-1}\llcorner \dd \, \Im \Omega^{(0)}, \qquad J_1 = J^{(0)} +\frac{1}{8\tilde{m}}J^{(0)-1}\llcorner \dd \, \Im \Omega^{(0)}\,.
\end{equation}

\section{Computation of the O6-backreaction in terms of one-forms}
\label{app:O6_localisation_one-forms}
The calculation of the O6-backreaction can be set up by making use of the three complex one-forms in which $J$ and $\Omega$ can be decomposed. Indeed, one can write them as follows:
\begin{equation}
    J = \frac{i}{2}\sum_{k=1}^{3} z_k \wedge \bar{z}_k, \qquad \Omega = z_1 \wedge z_2 \wedge z_3 \,.
\end{equation}
As explained in the main text, the anisotropy of the manifold makes the perturbative Ansatz rather complicated. For that reason, one needs to turn on second-order terms that eventually source some equations at first order. The one-forms can then be written as 
\begin{equation}
    z_k = z_k^{(0)} + z_k^{(1)} + z_k^{(2)}+ \cdots\,,
\end{equation}
where the various terms scale as
\begin{equation}
\label{eq:zcor_scalings}
    z_k^{(p)} \sim L_k n^{-p b} \,,
\end{equation}
with $L_k = (L_T, L_2, L_3)$.
The leading order pieces are the following complex one-forms:
\begin{equation}
    z_1^{(0)} = -L_T (e^6 + i e^1), \qquad z_2^{(0)} = L_2 (e^4 + i e^5), \qquad z_3^{(0)} = L_3 (-e^5 + i e^3) \,. 
\end{equation}
The corrections to the one-forms eventually give rise to the corrected SU(3)-structure forms $J$ and $\Omega$. At leading order, the two-form has components that scale differently, whereas the components of $\Omega$ scale all the same. As a consequence, one must write
\begin{align}
    &J^{(1)} = \frac{i}{2}\sum_{k=1}^{3} \left(z_k^{(0)} \wedge \bar{z}_k^{(1)} + z_k^{(1)} \wedge \bar{z}_k^{(0)} \right) + \frac{i}{2}\sum_{l=2,3} \left(z_l^{(0)}\wedge \bar{z}_l^{(2)} + z_l^{(2)}\wedge \bar{z}_l^{(0)} + z_l^{(1)}\wedge \bar{z}_l^{(1)}\right), \\
   & \Omega^{(1)} = \epsilon_{ijk} z^{(1)}_{i} \wedge z^{(0)}_{j} \wedge z^{(0)}_{k}, \qquad \Omega^{(2)} = \epsilon_{ijk} z^{(2)}_{i} \wedge z^{(0)}_{j} \wedge z^{(0)}_{k} + \epsilon_{ijk} z^{(1)}_{i} \wedge z^{(1)}_{j} \wedge z^{(0)}_{k}\,.
\end{align}
These corrections must solve
\begin{align}
    &F_2^\text{SU(3)} - F_2^\text{BI} = \mathcal{O}\left(n^{-b}\right)\,,\\
    &\dd J - 2 \tilde{m} \e^{-A}\Re \Omega = L_T^2 \; \mathcal{O}\left(n^{-2b}\right)\,,
\end{align}
which must be supplemented with the Bianchi identity
\begin{equation}
    \dd F_2^\text{BI} = -2 \sum_i\delta_{\text{O6}_i}\,.
\end{equation}
As explained in the main text, at first order, the intersecting O6-planes can be treated as a linear superposition of single sources. 
For that reason, it is sufficient to find corrections for one stack of O6-planes, as all the others are similar and can be summed up to the final result. Here, an O6-plane that wraps the three-cycle which has the Poincar\'e dual form $e^{123}$ is chosen, meaning that it generates a source $\delta e^{456}$ in the $F_2$ Bianchi identity.

As explained in Appendix D of ref.~\cite{Cribiori:2021djm}, one can neglect derivatives along the 1- and 6-direction, due to the anisotropy of the manifold. The associated length scale $L_T$ is much smaller than the others, $L_2$ and $L_3$, and it was shown that such derivatives generate exponentially small contributions in the flux parameter $n$.
This simplifies the problem. As usual, one should write the exterior derivative as 
\begin{equation}
    \dd = e^a \mathcal{D}_a, \qquad [\mathcal{D}_a , \mathcal{D}_b] = -f_{ab}^{c} \mathcal{D}_c\,,
\end{equation}
where $\mathcal{D}_a = e^\mu_a\partial_\mu$ and the $f^c_{ab}$ are the structure constants that were implicitly defined in this paper for the Iwasawa manifold in eq.~\eqref{eq:geomflux} as $\dd e^c = (f_{ab}^c /2) e^{ab}$. When the derivatives $\mathcal{D}_1$ and $\mathcal{D}_6$ can be ignored, one sees that all the derivatives commute and can be replaced by ordinary partial derivatives. 
With all this at hand, the corrections can be computed. The first-order corrections for the one-forms are then found to be the following:
\begin{align}
    &\Re z_1^{(1)} =g_s L_T\left(- \frac{\varphi_i}{L_T} e^6 -\alpha \frac{L_2}{L_3L_T}\partial_5\chi_i e^2+\beta\frac{L_3}{L_2L_T}\partial_4\chi_i e^3\right)\,,\\
    &\Im z_1^{(1)} = g_s L_T\left(\frac{\varphi_i}{L_T} e^1 +(\alpha + 3/2)\frac{L_2}{L_3L_T}\partial_5\chi_i e^4+(\beta - 3/2)\frac{L_3}{L_2L_T}\partial_4\chi_i e^5\right)\,,\\ 
    &\Re z_2^{(1)} = g_s L_2 \left(\frac{\varphi_i}{L_T}e^4\right)\,,\\
    &\Im z_2^{(1)} = g_s L_2 \left(-\frac{\varphi_i}{L_T}e^2\right)\,,\\
    &\Re z_3^{(1)} = g_s L_3 \left(-\frac{\varphi_i}{L_T}e^5\right)\,,\\
    &\Im z_3^{(1)} = g_s L_2 \left(-\frac{\varphi_i}{L_T}e^3 \right)\,.
\end{align}
and the relevant second-order pieces are 
\begin{align}
    &\Re z_2^{(2)} = g_s L_2 \left(\alpha 2 \tilde{m}\partial_5 \chi_i e^1+ \frac{5}{2} \tilde{m} \chi e^4\right)+ L_2 \left( \frac{g_s^2}{L_T^2}\varphi_i^2 (1-h_2)e^4\right) + \cdots\\
    &\Im z_2^{(2)} = g_s L_2 \left( -2 \tilde{m}(\alpha+5/2) \partial_5 \chi_i e^6 + \frac{5}{2}\tilde{m}\chi e^2 \right)+ L_2 \left( \frac{g_s^2}{L_T^2}\varphi_i^2 h_2 e^2\right) + \cdots\\
    &\Re z_3^{(2)} = g_s L_3 \left(-2 \tilde{m}\beta \partial_4 \chi_i e^1 -\frac{5}{2}\tilde{m} \chi e^5\right)+ L_3 \left( - \frac{g_s^2}{L_T^2}\varphi_i^2 (1-h_3)e^5\right) + \cdots\\
    &\Im z_3^{(2)} = g_s L_2 \left(2 \tilde{m}(\beta-5/2) \partial_4 \chi_i e^6 +\frac{5}{2}\tilde{m}\chi e^3\right)+ L_3 \left( \frac{g_s^2}{L_T^2}\varphi_i^2 h_3 e^3\right) + \cdots
\end{align}
Four parameters are unfixed, namely $\alpha$, $\beta$, $h_2$ and $h_3$. 
Reminding that $g_s/L_T \sim n^{-b}$ and $g_s 2\tilde{m} \sim n^{-2b}$, one verifies the scalings in eq.~\eqref{eq:zcor_scalings}.
Note that in the second-order corrections, the $\varphi_i^2$ terms are introduced to make sure that such terms do not appear in the correction of $J^{(1)}$, compensating the $(z^{(1)}_l \wedge \bar{z}_l^{(1)})$-terms.
In the presence of multiple O6-plane stacks, all contributions linear in the $\varphi_i$ and $\chi_i$ should be added, but the quadratic terms $\varphi_i^2 e^j$ should be replaced in $z_k$ by $(\sum_{i} \epsilon_{i,j} \varphi_i)^2 e^j$, where $\epsilon_{i,j} =+1$ if the $i$-th O6-plane is orthogonal to direction $j$ and $-1$ if parallel, such that $J^{(1)}$ is still linear in all the $\varphi_i$'s. 

One might conclude that all second-order corrections are irrelevant for $\Omega$ at the order the equations are being solved, as they would appear at second order there, but it is more subtle than that. Remember that in the SU(3)-structure expression of $F_2$, one has the contribution $J^{-1}\llcorner \dd (\e^{-3A}\Im\Omega)$. The $J^{-1}\llcorner$ operator introduces inverse lengths squared, improving the scaling of specific terms. Since the manifold is anisotropic and $J$ has components that scale differently, it depends on the legs of $ \dd (\e^{-3A}\Im\Omega)$ whether these second-order contributions appear at first order in $F_2$ or not.
With this in mind, the relevant corrections to $J$ and $\Omega$ can be computed and for $J$ it becomes
\begin{equation}
    J^{(1)} =  2\tilde{m} g_s \left( L_3^2\partial_4 \chi_i e^{56} - L_2^2 \partial_5 \chi_i e^{46} + \frac{5}{2}\chi_i (- L_2^2 e^{24}+L_3^2 e^{35})\right)\,.
\end{equation}
The relevant corrections to $\Im \Omega$ are summarised as follows:
\begin{align}
\label{eq:OmegariCorF1}  
     (e^{-3A}\Im \Omega)^{(1)} =& g_s L_T L_2 L_3 \frac{4\varphi_i}{L_T}\left(e^{256}+e^{346}+e^{145} \right) - g_s \frac{3}{2}\left(L_2^2 \partial_5 \chi_i e^{234}-L_3^2\partial_4 \chi_i e^{235}\right)\,,\\  
\label{eq:OmegariCorF2} 
(e^{-3A}\Im \Omega)^{(2)} =& 
- g_sL_T^2\frac{5}{2} \left(-\partial_4 \chi_i e^{126}+\partial_5 \chi_i e^{136} \right) + \cdots
\end{align}
and for $\Re \Omega$ it is found that
\begin{align}
    (e^{-A}\Re \Omega)^{(1)} =& g_s L_T L_2 L_3 \left(\frac{4\varphi_i}{L_T} e^{456} \right) - g_s \frac{3}{2}\left(L_2^2 \partial_5 \chi_i e^{245}+L_3^2\partial_4 \chi_i e^{345}\right)\,,\\
    (e^{-A}\Re \Omega)^{(2)} =&- g_s L_T^2\frac{5}{2}\left(\partial_5 \chi_i e^{156}+\partial_4 \chi_i e^{146}\right) + \cdots
\end{align}
Note that the only form that receives relevant corrections in $\chi_i$ without derivatives is $J^{(1)}$.
The corrections can also be written in terms of the zeroth order forms $J^{(0)}$ and $\Omega^{(0)}$ and also in terms of $F_2^{(0)}$, which is how the results were presented in the main text. For $J$ one finds
\begin{equation}
    J^{(1)} = g_s J^{(0)-1}\llcorner \left( \dd \chi_i \wedge( \Im \Omega^{(0)}- k_{||,i})\right) -  \frac{5}{8}\: g_s \chi_i J^{{(0)}-1}\llcorner \dd \, \Im \Omega^{(0)}\,,
\end{equation}
whereas the corrections of $\Omega$ take the following form:
\begin{align}
    (\e^{-3A}\Im\Omega)^{(1)}+(\e^{-3A}\Im\Omega)^{(2)} &= \frac{g_s\varphi_i}{L_T} ( \Im \Omega^{(0)}- k_{||,i}) -I^{(0)} \cdot \left(g_s\dd \chi_i \wedge F_2^{(0)} \right)\,,\\
    (\e^{-A}\Re \Omega)^{(1)}+(\e^{-A}\Re \Omega)^{(2)} &= \frac{4 g_s \varphi_i}{L_T} k_{\perp,i} + g_s \dd \chi_i \wedge F_{2}^{(0)}\,.
\end{align}

\subsubsection*{Subleading corrections at some places become leading in others}
Some aspects of the leading vs. subleading corrections are now clarified.
First, in eq.~\eqref{eq:OmegariCorF2}, only the second-order corrections that become first order in other $F_2^\text{SU(3)}$ are kept. To illustrate how that works, consider the exterior derivative on this equation (which is nothing more than 
$\frac{5}{2} g_s \mathcal{C}_1 \wedge J_1)$, which would generate the following terms: 
\begin{align}
\label{eq:scalingdemonstration}
\begin{split}
    g_sL_T^2\frac{5}{2}\, \dd  \left(-\partial_4 \chi_i e^{126}+\partial_5 \chi_i e^{136} \right) =&  g_sL_T^2\frac{5}{2}\, \left(-\partial_4^2 \chi_i e^{1624}+\partial_5^2 \chi_i e^{1635} \right)\\
    &- 2\tilde{m} g_s \frac{5}{2} (\partial_4 \chi_i e^2 - \partial_5 \chi_i e^3) \wedge \Re \Omega^{(0)}\,.
\end{split}
\end{align}
If one now acts with $g_s^{-1}J^{(0)-1}\llcorner$ on the first term, one gets the following:
\begin{align}
\begin{split}
    L_T^2\frac{5}{2}\, J^{(0)-1} \llcorner \left(-\partial_4^2 \chi_i e^{1624}+\partial_5^2 \chi_i e^{1635} \right) =& L_T^2\left(\frac{\partial_4^2}{L_2^2} +\frac{\partial_5^2}{L_3^2}\right)\chi_i  e^{16} + \left(\partial_4^2 \chi_i e^{24}-\partial_5^2 \chi_i e^{35} \right)\\
    =& L_T^2 8\tilde{m}  \frac{\varphi_i}{L_T}e^{16} +\left(\partial_4^2 \chi_i e^{24}-\partial_5^2 \chi_i e^{35} \right)\,.
\end{split}
\end{align}
The first term scales as $n^{-b}$, and hence contributes only at subleading order in the Bianchi identity of $F_2$. In contrast, the last term does not scale and contributes at leading order in the Bianchi identity for $F_2$. 
Writing out the second term of eq.~\eqref{eq:scalingdemonstration} a bit more, one gets
\begin{align}
\begin{split}
    &-5\tilde{m}g_s \left((\partial_4 \chi_i e^2 - \partial_5 \chi_i e^3) \wedge \Re \Omega^{(0)} \right) =  5\tilde{m} g_s \left( \dd \chi_i \wedge \Im \Omega^{(0)}\right)\\
    & =5 L_T L_2 L_3\tilde{m} g_s \left[ -\partial_4 \chi_i\left( e^{1234}+ e^{2456}\right) -\partial_5 \chi_i\left( e^{1235}- e^{3456}\right) \right]\,,
\end{split}
\end{align}
and acting with $g_s^{-1}J^{(0)-1}\llcorner$, one finds
\begin{equation}
     5\tilde{m} J^{(0)-1}\llcorner \left(\dd \chi \wedge \Im \Omega^{(0)} \right) = -\frac{5}{2} L_T^2 \left[ \frac{\partial_4 \chi_i}{L_2^2}\left( e^{13}-e^{56}\right) +\frac{\partial_5 \chi_i}{L_3^2}\left( e^{12} + e^{46}\right) \right]\,.
\end{equation}
This indeed scales as $n^{-b}$ and hence is subleading in $F_2^\text{SU(3)}$. It turns out that the second-order terms that are included, are sufficient and that others will not source $F_2^\text{SU(3)}$ at leading order.

At last, it also has to be emphasised that the corrections in $J$ also have to be taken into account in the operator $J^{-1}$. By using the definition \eqref{eq:Jcontract_Idot}, one sees that it satisfies
\begin{equation}
    J^{-1}\llcorner \;= J^{(0)-1}\llcorner\; - (J^{(0)-1}\cdot J^{(1)} \cdot J^{(0)-1})^{-1}\llcorner \,.
\end{equation}
However, the only term that the first order correction of $J^{-1}\llcorner$ can act on, is $\dd \, \Im \Omega^{(0)}$, which results in subleading corrections only (i.e. $\mathcal{O}(n^{-b})$). In fact, one finds
\begin{equation}
    (J^{(0)-1}\cdot J^{(1)} \cdot J^{(0)-1})^{-1}\llcorner \dd \, \Im \Omega^{(0)}= -5 \tilde{m} \chi_i J^{(0)-1} \dd \, \Im \Omega^{(0)},
\end{equation}
which scales as $n^{-b}$ indeed.

\subsection*{The metric}
\label{app:metric}
Once $\Omega$ and $J$ are found, the metric of the 6d space can be constructed, by first computing the corrected almost complex structure $I$ an almost symplectic structure $J$ and using
\begin{equation}
    \dd s_6^2 = g_{ab}e^a  e^b = J_{al}I^l_b e^a e^b\,.
\end{equation}
Alternatively, one can take the one-forms and write
\begin{equation}
    g_{ab}e^a e^b = \sum_{k=1}^{3} z_k \; \bar{z}_k.
\end{equation}
This results in the following metric:
\begin{align}
    \dd s_6^2 =& \left(1+2\frac{g_s\varphi_i}{L_T}\right) \left[(L_2 e^4)^2+(L_3 e^5)^2+(L_T e^6)^2 \right]\\
    &+\left(1-2\frac{g_s\varphi_i}{L_T}\right) \left[(L_2 e^2)^2+(L_3 e^3)^2+(L_T e^1)^2 \right]\\
    &-5 \left(g_s\frac{ \partial_5\chi_i}{L_3}(L_2 e^2)+ g_s\frac{ \partial_4\chi_i}{L_2}(L_3 e^3)\right)(L_T e^{6})\\
    &-3 \left(g_s\frac{ \partial_5\chi_i}{L_3}(L_2 e^4)-g_s \frac{ \partial_4\chi_i}{L_2}(L_3 e^5)\right)(L_T e^{1})\,.
\end{align}
Here, metric terms at order $\mathcal{O}(n^{-2b})$ have been discarded. However, one sees that off-diagonal terms survive at order $\mathcal{O}(n^{-3b/2})$, as $g_s/L_2 \sim g_s/L_3 \sim n^{-3b/2}$.
The diagonal components at order $\mathcal{O}(n^{-b})$ match with the metric computed in ref.~\cite{Cribiori:2021djm}. The off-diagonal components at lower order did not appear there. However, it turns out that they do not contribute at first order in the 10d equations of motion and hence do not alter the results of ref.~\cite{Cribiori:2021djm}. 
Instead, these corrections generate additional off-diagonal components of the Ricci tensor. Indeed, to the lowest non-trivial order, there are four off-diagonal components, $R_{14}$, $R_{15}$, $R_{26}$ and $R_{36}$. They are generated both by the diagonal terms in the metric with the $\varphi_i$'s and by the non-diagonal terms with the $\chi_i$'s. As it turns out, the latter is crucial for solving the 10d Einstein equations, as is shown now.
In ref.~\cite{Cribiori:2021djm}, all the Einstein equations were written down and their diagonal components were solved at next-to-leading order.
The lowest-order off-diagonal components of the Einstein equations, adapted to the notation here, are
\begin{equation}
    R^{(1)}_{mn} -4 \nabla_m \nabla_n \e^{A} - 2 g_s \nabla_m \nabla_n \e^{-\phi} = \frac{1}{2}g_s^2|F_2|^2_{mn}\,, \qquad m\neq n \,.
\end{equation}
As an example, take $R_{14}$. The diagonal terms of the metric generate a contribution with a single derivative on $\varphi_i$, as the underlying torus is twisted. This results in the following contribution to the lowest order:
\begin{equation}
    R^{(\varphi_i)}_{14} = 6\tilde{m} g_s \frac{L_2}{L_3} \partial_5 \varphi_i \,.
\end{equation}
The contribution from the off-diagonal terms is computed to be
\begin{equation}
    R^{(\chi_i)}_{14} = 6\tilde{m} g_s \frac{L_2}{L_3} \partial_5 \varphi_i \,.
\end{equation}
Additionally, for the corrections for the warp factor and dilaton, one notices that the relevant terms appear from $\nabla_m \nabla_n \supset -\Gamma^k_{mn} \partial_k$, and results in
\begin{align}
    &-4 \nabla_1 \nabla_4 \e^{A} = + 4 \;\Gamma^n_{14}\;\partial_n \e^{A} = 4\tilde{m} g_s \frac{L_2}{L_3} \partial_5 \varphi_i\,,\\
    &- 2 g_s \nabla_1 \nabla_4 \e^{-\phi} = + 2 g_s \;\Gamma^n_{14}\;\partial_n \e^{-\phi} = -6\tilde{m} g_s \frac{L_2}{L_3} \partial_5 \varphi_i\,.
\end{align}
Everything combined, one finds:
\begin{equation}
 R^{(1)}_{14} -4 \nabla_1 \nabla_4 \e^{A} - 2 g_s \nabla_1 \nabla_4 \e^{-\phi} = 10 \tilde{m} g_s \frac{L_2}{L_3} \partial_5 \varphi_i\,.
\end{equation}
This corresponds indeed to
\begin{equation}
    \frac{1}{2}g_s^2|F_2|^2_{14} = \frac{1}{2}g_s^2 F_{16}^{(0)}F_{46}^{(1)} g^{(0)66} = 10 \tilde{m} g_s \frac{L_2}{L_3} \partial_5 \varphi_i\,.
\end{equation}
For the 15-, 26- and 36-components the same match appears. Hence, the off-diagonal terms of the metric that were found by localising the scale-separated model with pure spinor techniques, can solve off-diagonal components of the Einstein equations at the lowest non-trivial order, which is a nice achievement.
\bibliographystyle{JHEP}
\bibliography{AllRefs.bib}

\end{document}